\documentclass[12pt]{article}
\usepackage{bm}
\usepackage{latexsym}
\usepackage{dcolumn}
\usepackage{amsfonts,amssymb}
\usepackage{graphicx,epsfig}
\usepackage{psfrag}
\usepackage{amsmath,amssymb}
\def\be{\begin{equation}}
\def\ee{\end{equation}\noindent}
\def\ben{\begin{equation}}
\def\een{\end{equation}\noindent}
\def\ba{\begin{eqnarray}}
\def\ea{\end{eqnarray} \noindent}
\def\bea{\begin{eqnarray}}
\def\eea{\end{eqnarray} \noindent}
\def\nn{\nonumber}

\def\mcR{\mathcal{R}}

\def\sech{\text{sech}}
\def\TT{\mathcal{T}}
\def\bz{{\bar{z}}}
\def\tq{{\tilde{q}}}

\begin{document}
\begin{titlepage}
\thispagestyle{empty}
\begin{flushright}
UK/10-11
\end{flushright}

\bigskip

\begin{center}
\noindent{\Large \textbf
{On Dumb Holes and their Gravity Duals}}\\
\vspace{2cm} \noindent{
Sumit R. Das\footnote{e-mail:das@pa.uky.edu},
Archisman Ghosh \footnote{e-mail:archisman.ghosh@uky.edu},
Jae-Hyuk Oh\footnote{e-mail:jaehyukoh@uky.edu} $~$and
Alfred D. Shapere \footnote{e-mail:shapere@pa.uky.edu}}

\vspace{1cm}
  {\it
 Department of Physics and Astronomy, \\
 University of Kentucky, Lexington, KY 40506, USA\\
 }
\end{center}

\vspace{0.3cm}
\begin{abstract}
Inhomogeneous fluid flows which become supersonic are known to produce
acoustic analogs of ergoregions and horizons.
This leads to Hawking-like radiation of phonons with a
temperature essentially given by the gradient of the velocity at the
horizon. We find such acoustic dumb holes in charged conformal fluids and use
the fluid-gravity correspondence to construct dual gravity
solutions. A class of quasinormal modes around these gravitational
backgrounds perceive a horizon. Upon quantization, this implies a
thermal spectrum for these modes.
\end{abstract}
\end{titlepage}
\newpage

\tableofcontents
\newpage

\section{Introduction}
\label{intro}

In recent years, gauge-string duality
\cite{Maldacena:1997re,Aharony:1999ti} has been useful in exploring
properties of strongly coupled field theories in regimes where
their duals may be truncated to classical gravity. In particular,
application of gauge-string duality to the hydrodynamic regime of these field
theories has led to a ``fluid-gravity correspondence''
\cite{Policastro:2001yc}-\cite{Bhattacharyya:2008ji}. Properties of solutions of classical gravity
then lead to predictions for interesting properties of the dual fluid, 
the most celebrated example being the ratio of shear viscosity to
the entropy density of a conformal fluid \cite{Policastro:2001yc}.

In this note we use the fluid--gravity correspondence in the opposite
direction. We use properties of supersonic fluid flows to predict
interesting properties of fluctuations around a class of deformed black
brane spacetimes in asymptotically $AdS$ spacetimes.  These
spacetimes are duals of inhomogeneous flows of conformal fluids
where the fluid velocity exceeds the speed of sound in some region. 
Unruh \cite{Unruh:1980cg}
showed that such flows lead to the formation of an
``acoustic ergoregion'' and, under suitable conditions, to an
``acoustic horizon''. The same physics which leads to Hawking
radiation from black holes in General Relativity now leads to a
Hawking-like radiation of quantized sound waves (or phonons) with a
thermal spectrum, the temperature being proportional to the gradient of the
velocity field at the acoustic horizon
\cite{Visser:1993ub,Barcelo:2000tg}.  Even when an acoustic horizon
is not present, the presence of an ergoregion leads to
characteristic properties like superradiance
\cite{Basak:2002aw} .  Fluid configurations
with such acoustic horizons have been termed ``dumb holes'', and
have been proposed as possible experimentally realizable systems 
for testing the physics of Hawking radiation in the laboratory
\cite{reviews} . 

We will show that the gravity duals of such supersonic flows are
non-static black holes. The duals of sound waves are then certain 
quasinormal
modes around such black holes, and it follows from the fluid-gravity
correspondence that at the quantum level one should find a 
Hawking-like radiation of these modes with an approximately
thermal spectrum \cite{Policastro:2002tn}. 
This Hawking-like radiation is distinct
from the usual Hawking radiation associated with the black hole
horizon, and would be present even when the background black hole is
extremal and hence at zero temperature. The temperature of this
quasinormal mode radiation depends on the properties of a ``quasinormal
mode horizon'', which is an extension  into the bulk 
of the acoustic horizon of the boundary fluid.

It should be emphasized that this phenomenon could have been found
purely in General Relativity (or its supergravity extensions relevant
to our considerations) by studying the fluctuation problem around
these non-static black holes. However, without the fluid gravity
correspondence and knowledge of acoustic Hawking radiation, there
would not have been an obvious motivation to look for quasinormal mode horizons in 
non-static black brane backgrounds.

While we believe that the phenomenon of Hawking radiation or
super-radiance of quasinormal modes is quite general, it turns out to
be rather difficult to come up with examples within a controlled
approximation scheme.  The simplest and perhaps most interesting
background where such a phenomenon could be present is a Kerr black
hole in asymptotically $AdS_5$ spacetime. The dual of such a background 
is a rotating conformal fluid on $S^3$ 
\cite{Awad:1999xx,Bhattacharyya:2007vs}. 
There is a regime of parameters of the black hole geometry for which the
dual rotating fluid has supersonic velocities in a band around the
equator of the $S^3$, thus producing an ergoregion for sound
modes. The physics of sound waves around such a rotating fluid
background would have a dual description in terms of quasinormal modes
of gravitational perturbations around the Kerr black hole in $AdS_5$.
However, this flow has vorticity, and the existence of acoustic Hawking
radiation has been demonstrated mostly for irrotational flow. In the
presence of nonzero vorticity, the sound modes 
get mixed up badly with other modes and analysis becomes difficult
\cite{PerezBergliaffa:2001nd}.

The situation simplifies, however, if the flow is irrotational.  As
shown in \cite{Unruh:1980cg} for non-relativistic perfect fluids, and
extended to relativistic perfect fluids in \cite{Bilic:1999sq}, the velocity
potential then obeys the wave equation for a minimally coupled massless
scalar field propagating on a curved background, the metric of which
is determined by the underlying flow.
The mathematical problem of quantizing sound waves or phonons around
such a flow is then quite similar to that
of quantizing a massless scalar field in an ordinary black
hole background. This implies the existence of an acoustic analog of Hawking
radiation. 

Known examples of supersonic flows of perfect fluids often lead to
infinite ``acoustic surface gravity'' (which is proportional to the
gradient of the velocity at the acoustic horizon). The presence of
viscosity usually regulates this divergence and renders it finite
\cite{Liberati:2000pt}. The incorporation of viscosity, however, makes the
analysis complicated.

In this paper we find simple examples of acoustic horizons in
{\em ideal} relativistic conformal fluids with finite acoustic
Hawking temperature. The simplest example involves a fluid moving in
a background spacetime of the form
\ben
ds_B^2 = -dt^2 + dz^2 + R(z)^2(d\theta^2 + \sin^2\theta d\phi^2)
\label{intro-1}
\een
where $R(z)$ is a slowly varying function which has the behavior $R(z)
\rightarrow |z|$ as $|z| \rightarrow \infty$.\footnote{The precise meaning of a slowly-varying $R(z)$ is given in the discussion above Eqn.(\ref{deriv_curv}). $R(z)$ is some function in the metric (\ref{intro-1}), for which all invariants constructed out of the curvature and its derivatives are small compared to the basic scale in the problem.}
The fluid flow is steady
and the only nonzero
component of the fluid velocity is $v_z(z)$, with all
derivatives bounded. Starting with $v_z = 0$ at $z= -\infty$ we will
show that $v_z$ reaches the speed of sound -- producing an acoustic horizon -- at
minima of $R(z)$.   If the function $R(z)$ has only one minimum,
{\it e.g.} $R(z) = \sqrt{(z^2+z_0^2)}$, the assumption of a smooth solution of the fluid
equations of motion implies that the fluid velocity continues to
increase beyond the acoustic horizon until it reaches the speed of light
at $z = \infty$. However, if $R(z)$ has multiple extrema, {\it e.g.} two
minima separated by a maximum, we will find smooth flows where the
fluid velocity reaches a maximum supersonic value at the maximum of
$R(z)$ and then decreases, then turns subsonic at the second minimum,
and finally reaches zero at $z = \infty$. Sound waves cannot escape 
to the asymptotic region $z=-\infty$ from beyond the first acoustic horizon
(which is therefore like a black hole horizon), and cannot cross the second acoustic horizon 
(which behaves more like a white hole horizon) from the $z=\infty$ asymptotic region .

We also study flows in a warped $R^{1,1} \times T^2$ geometry
\ben
ds^2 = -dt^2 + R(z)^2 (d\theta_1^2+d\theta_2^2) + dz^2
\label{intro-3}
\een
and find very similar phenomena.

The acoustic Hawking radiation that arises when the sound modes of these
flows are quantized will be easiest to detect if the acoustic Hawking 
temperature $T_H$ is larger than the ambient
temperature of the fluid, $T_H > T$. For an uncharged conformal fluid,
the only scale is the temperature, so that hydrodynamics is valid
only when all derivatives are small compared to the
temperature. However $T_H$ itself is proportional to the gradient of
the velocity field, $T_H \sim \frac{dv_z}{dz} |_{z=\bz}$. 
Thus it is not possible to have Hawking radiation at a temperature higher than the ambient temperature in an uncharged conformal fluid.
We will therefore look for {\it charged} fluid solutions satisfying the rather stringent criterion of $T_H>T$, in order to ensure detectability of Hawking radiation.\footnote{We would like to thank the referee and also Dileep Jatkar for bringing our attention to the work of Weinfurtner {\em et.al.\!} \cite{Weinfurtner:2010nu} where analogue Hawking radiation at a very low temperature has been observed in a background that has a temperature several orders of magnitudes higher than the radiation.}  
Note that the solutions we will find all have well-defined uncharged limits and satisfy all criteria other than $T_H>T$ under the replacement of $\TT\to T$ and $q\rightarrow0$.

We will consider conformal fluids with global charge density $q$ and conjugate chemical potential $\mu$.  For such a
fluid in equilibrium, 
the ambient temperature can be made to vanish provided the
charge density is chosen properly.  In analogy with the gravitational
duals of such fluids considered below, we will call such a fluid
``extremal''. We will consider 
{\em
isentropic} flows of such a fluid where the charge density $q$, energy
density $\epsilon$ and the velocity vary slowly (in a sense defined
precisely below) but whose variations can themselves be $O(1)$. 
In an isentropic flow, the entropy density per unit charge density 
is, however,  a constant. 
For such
flows $q \propto \epsilon^{3/4}$ where $\epsilon$ is the energy
density.
Then $\TT\propto\epsilon^\frac{1}{4}$ is the only independent energy
scale in the theory.  $\TT$ is in general a function of the chemical
potential $\mu$ and the temperature $T$.
As shown below, the isentropic condition allows us to keep the local
temperature $T$ to be always much smaller than $\TT$, even though the
other hydrodynamic quantities can change by $O(1)$.
In the limit of very small temperature $\TT \propto \mu$.  One
would expect that the hydrodynamic approximation is valid so long as
all gradients are small compared to $\TT$. We will show that it is
consistently possible to construct fluid flows described above with
all gradients $\frac{dv_z}{dz}$ and all curvature invariants 
much smaller than $\TT$, thus ensuring $\TT \gg T_H \gg T$.

The spacetimes (\ref{intro-1}) and (\ref{intro-3}) may each be
regarded as the boundary of an
asymptotically $AdS_5$ near-extremal charged black brane geometry
which is deformed due to a nonzero boundary curvature. 
For generic spacetimes with arbitrary $R(z)$, the boundary metric might not admit a smooth bulk dual.
However it has been shown in \cite{Bhattacharyya:2008ji} that for slow-enough variations in the boundary metric,
the dual geometry is regular up to the bulk horizon and free of other singularities.
Our solution will admit a smooth bulk dual as long as all invariants constructed out of the boundary curvature and its derivatives are 
much smaller than the radius of the outer horizon $R_+$ in $AdS$ units.
One should note that in spite of being slowly-varying, the deformations can still be large. 
Very close to extremality, $R_+ \sim \TT \sim \mu$, so that
this condition is in fact the condition for validity of hydrodynamics
in the boundary theory. In the second example, the boundary metric has
two compact directions. This means that the nature of the dual
geometry depends on the size of the compact directions compared to
$R_+$ \cite{Nishioka:2009zj}. 
We will choose $R(z) \gg R_+$ so that the dual is a
near-extremal black brane rather than an $AdS$ soliton with a small
temperature. 

A fluid flow profile in the boundary theory is then described by a
normalizable deformation of this bulk metric. 
We construct the
deformed bulk metric using a derivative expansion,
following \cite{Bhattacharyya:2008jc}, 
\cite{Erdmenger:2008rm} and \cite{Banerjee:2008th}. The
straightforward derivative expansion breaks down in ``tubes'' of
constant retarded time
where the geometry becomes exactly extremal; 
therefore, we consider fluid flows where the local temperature is small
but nonzero. We then consider the class of linearized fluctuations 
around this background geometry - 
quasinormal modes - which are dual to sound waves in the presence of the
corresponding fluid flow.  Note that while the deformations of the
bulk metric due to a nontrivial $R(z)$ and a nontrivial velocity
profile $v_z(z)$ are typically large, the quasinormal mode amplitudes
are small.

The behavior of the quasinormal modes clearly shows that at leading
order in the derivative expansion, the acoustic horizon of the fluid
extends into the bulk in the following sense. Let $r$ denote the
radial coordinate in the $AdS$ space and let $z=\bz$ be the location
of the acoustic horizon in the boundary flow.  We find that for any
value of $r$, these quasinormal modes suffer an infinite blue-shift as
we approach $z = \bz$: modes which travel along the direction of the
fluid flow are smooth at $z= \bz$, while the modes which travel in the
direction opposite to the flow have rapid oscillations. Thus, in an
eikonal approximation these quasinormal modes cannot cross the
quasinormal mode horizon at $z=\bz$, which extends radially from the
acoustic horizon into the bulk.

Standard arguments imply that upon quantization\footnote{Quantization
of bulk modes corresponds to $1/N$ corrections in the $SU(N)$ gauge
theory on the boundary.}, one would find a thermal distribution of
these quasinormal modes with a temperature $T_H$, which is the gravity
dual of the acoustic Hawking radiation in the fluid. Only these
specific quasinormal modes perceive the quasinormal mode horizon;
other modes can cross it with ease. By the same token, the thermal
distribution will be made up only of these quasinormal modes; it
exists independently of (and at a different temperature from) the
usual Hawking radiation associated with the event horizon of the
background black brane.

Our discussion is restricted to the lowest non-trivial order in the
derivative expansion, which is consistent with the perfect fluid
approximation. However we expect that the physical consequences should
survive higher derivative corrections. Furthermore our discussion of
{\em fluctuations}, both in the boundary fluid and in bulk gravity, is
restricted to the linearized limit. We do not address the effect of
nonlinear interactions of the sound waves and other modes.

Admittedly, our setup is a bit contrived and is meant to provide a
simple toy model in which this novel gravitational phenomenon can be
studied in a controlled fashion.  We expect, however, that the
phenomenon is quite general and would be present in more interesting
situations ({\it e.g.} the Kerr black hole mentioned above).

The paper is organized as follows. In Section \ref{acoustic_rel},
we give a self-contained discussion of acoustic metrics and dumb holes
for conformal relativistic fluids. In Section \ref{bulk}, we
describe the bulk dual. In Section \ref{regime_validity} we discuss
the regime of validity of our solutions.

\section{Acoustic metric for relativistic conformal fluid}
\label{acoustic_rel}

In this section we derive the equation governing the propagation of
sound around gradient flows of a perfect relativistic conformal fluid.
For such fluids, the pressure $p$ and the energy density $\epsilon$ are related by
\ben
p = \frac{\epsilon}{3}.
\label{defconformal}
\een
For a charged conformal fluid with charge density $q$, there is an additional equation of state $\epsilon=\epsilon(s,q)$, or equivalently $s=s(\epsilon,q)$,
that relates the energy, entropy and charge densities.~\footnote{For an uncharged conformal fluid, such a relation is trivial and $s\sim\epsilon^\frac{4}{3}$.}
We will eventually be considering charged fluids with gravity duals, and will write down an explicit equation of state for such fluids in Section (\ref{bulk}).

The first law of thermodynamics reads
\be
\label{1stlaw}
d\epsilon=T\,ds+\mu\,dq\,,
\ee
where $T$ and $\mu$ are the intensive quantities temperature and chemical potential respectively and can be obtained from the equation of state by taking derivatives:
\be
\label{T_mu}
T=\left.\frac{\partial \epsilon(s,q)}{\partial s}\right|_{q},\hspace{1cm}
\mu=\left.\frac{\partial \epsilon(s,q)}{\partial q}\right|_{s}.
\ee
For a homogeneous system, it follows from extensivity that all thermodynamic variables are related by a Gibbs-Duhem relationship, 
which using  (\ref{defconformal}) may be written as
\ben
\hbox{$\frac{4}{3}$}\epsilon = Ts+\mu q\,.
\label{gibbsduhem}
\een 

In the following, we will define a quantity $\TT$ with dimensions of energy by
\ben
 p = \frac{\epsilon}{3} = c\TT^4\,,
\label{ttandrr}
\een
where $c$ is a dimensionless constants depending on the underlying system. $\TT$ will be a function of $T$ and $\mu$ (or $q$), which reduces to $T$ in the uncharged limit.  Our fluids will also admit a zero-temperature, finite-$\mu$ limit, close to which $\TT$ is proportional to $\mu$.  $\TT$ sets the energy scale of our conformal fluid, and will play an  important role in defining limits in which our approximations are valid.

The equations of motion of fluid
dynamics are conservation of the energy momentum tensor and 
conservation of the currents associated with any conserved charges,
including the conserved particle number:
\begin{align} & \nabla_\mu T^{\mu\nu}
= 0 \nn \\ {\rm and} \hspace{2em}
& \nabla_\mu j_i^\mu = 0\,.
\label{fluid_eom}
\end{align}
The stress tensor $T^{\mu\nu}$ depends on the $3$
independent components of velocity ${\bf{v}}(x^\mu)$, the energy density, the pressure and their derivatives.
The currents $j_i^\mu$ additionally depend
on the densities $q_i$ of the conserved charges.
To leading order in the derivative expansion,
\ba
\label{Tmunu} 
T^{\mu\nu} &=& p\,g^{\mu\nu} + (\epsilon+p)u^\mu u^\nu = c\TT^4 \left( g^{\mu\nu} + 4u^\mu u^\nu \right) \\
\label{current}
j_i^\mu &=& q_i u^\mu\,. \ea
Here $u^\mu\equiv(\gamma,\gamma{\bf{v}})$ and
$\gamma=\frac{1}{\sqrt{1-v^2}}$.
Conformal invariance implies that the stress tensor is traceless. 
In a general curved background
there is a trace anomaly; however this is a higher order effect in
the derivative expansion and we have ignored it.
We have also ignored the viscous and
diffusive terms -- which are again higher order in the derivative
expansion -- and we therefore work in the perfect
fluid limit.  In addition we will also restrict attention to the case of a single charge of 
density $q(x^\mu)$.

The parallel  component of the equations of motion 
(\ref{fluid_eom}), $u_\nu\nabla_\mu T^{\mu\nu}=0$, leads to the conservation law \be
\label{entropy}
\nabla_\mu\left(\TT^3u^\mu\right) = 0\,. \ee 
In the uncharged case $\TT^3$
is proportional to the entropy density of the fluid
and the above equation is the conservation of the entropy current.
The perpendicular component $P^\lambda_\nu\nabla_\mu T^{\mu\nu}=0$
(where the projector $P^\lambda_\nu \equiv \delta^\lambda_\nu +
u^\lambda u_\nu$) gives 
\be
\label{perp}
u^\mu\nabla_\mu(\TT u^\nu)=-\nabla^\nu \TT
\ee
which can be manipulated to yield
\be
\label{perp_vort}
\nabla_\mu(\TT u_\nu) -\nabla_\nu(\TT u_\mu) = -\TT\omega_{\mu\nu}\,,
\ee
where $\omega_{\mu\nu} \equiv P_\mu^\lambda P_\nu^\kappa \left(
\partial_\lambda u_\kappa - \partial_\kappa u_\lambda \right) =0$ is
the vorticity of the fluid.\footnote{It can be shown that the condition of vanishing vorticity
is identical to the condition $\partial_\mu(fu_\nu) =
\partial_\nu(fu_\mu)$ for any scalar function $f$. In
(\ref{potential}) $f$ is simply the temperature $T$.}  Therefore, for
an irrotational flow, we can define a potential $\phi$ such that 
\be
\label{potential}
\TT u_\mu = \partial_\mu\phi\,. \ee
Thus to solve for irrotational flows of an uncharged fluid, it is 
sufficient to solve (\ref{potential}) and (\ref{entropy}), along with 
 an additional equation like
(\ref{current}) for every conserved charge. Note that since $u^\mu
u_\mu = -1$, the equation
(\ref{potential}) may be used to express $\TT$ in terms of the potential
$\phi$
\ben
\TT^2 = - (\partial_\mu \phi)(\partial^\mu \phi)
\label{TTform}
\een
so that $\phi$ determines both $u^\mu$ and $\TT$.

In general, the charge density $q$ is not related to $\TT$. We will,
however, restrict ourselves to solutions where $q/\TT^3$ is a
constant.
The current conservation equations
(\ref{fluid_eom}) are then automatically solved once the equation
(\ref{entropy}) is solved. For such flows $\TT(x)$ is the only independent dimensionful quantity that governs the flow.
$\phi(x)$ determines all the
hydrodynamic quantities once the ratio $q/\TT^3$ is specified.
In fact, substituting (\ref{TTform}) in (\ref{potential})
and finally in (\ref{entropy}) one gets a single complicated nonlinear
differential equation for $\phi(x^\mu)$.

Although the restriction that $q\sim\TT^3$ might seem quite ad hoc at this stage, for fluids with gravity duals that we will be considering in Section (\ref{bulk}), 
this will turn out to imply that the flow is {\em isentropic}.
Isentropic flows also allow us to parametrically control the
temperature of the fluid. As discussed above, we need to consider
fluids at low temperatures. In the flows we consider, derivatives of
the velocity, entropy etc. are small, even though their values can and
should change by $O(1)$. 
The equation (\ref{tmurelations}) shows
that once we fix the ratio $q/s$ so that $T$ is small at some time, it
remains parametrically small at all times, since the change of $s$ is
of order 1.

Isentropic sound waves in such a gradient flow 
are described by small amplitude fluctuations
of the velocity potential $\phi\to\phi+\delta\phi$. 
This induces variations of
$u^\mu, \TT$ and $q^i$, 
 $\TT\to\TT+\delta\TT,\
u^\mu\to{u^\mu}+\delta{u^\mu}, q_i \to q_i + \delta q_i$.  Plugging
these into the equations of motion (\ref{potential}), we get \ba
(\TT+\delta{\TT})(u^\mu+\delta{u^\mu}) &=& \partial^\mu\phi +
\partial^\mu\delta\phi \nn \\ {\rm or} \hspace{1cm} u^\mu\delta\TT +
\TT\delta u^\mu &=& \partial^\mu\delta\phi\,.
\label{variation}
\ea
Using $u_\mu\delta u^\mu=0$ we get
\ba
&& \delta\TT = -u^\mu\partial_\mu\delta\phi \nn \\
{\rm and} && \TT\delta u^\mu = 
P^{\mu\nu}\partial_\nu(\delta\phi)\,.
\label{lin_T_u}
\ea
Plugging (\ref{lin_T_u}) in (\ref{entropy})
\ba
\label{wave_eqn}
&& \nabla_\mu \left[ \left( 3\TT^2\delta\TT u^\mu +\TT^3\delta u^\mu
\right)\right] = 0 \nn \\ \implies && \partial_\mu
\left[\sqrt{-g}\TT^2\left( g^{\mu\nu}-2u^\mu u^\nu
\right)\partial_\nu\right](\delta\phi) =0 \ea For sound waves
in a static equilibrium fluid in flat spacetime,
this gives
$(-3\partial_t^2+\partial_i^2)(\delta\phi)=0$, from which we can read
off the speed of sound $c_s=\frac{1}{\sqrt{3}}$.  

More generally
(\ref{wave_eqn}) is the Klein-Gordon equation of motion of a 
{\em massless} scalar field in a non-trivial background metric,
\ben
\partial_\mu\left[\sqrt{-G}G^{\mu\nu}\partial_\nu\right](\delta\phi)=0
\een
where
 \ba
\sqrt{-G}G^{\mu\nu} &=& \sqrt{-g}\TT^2\left( g^{\mu\nu}-2u^\mu u^\nu \right) \nn \\
G_{\mu\nu} &=& \sqrt{3}\TT^2\left( g_{\mu\nu}+\frac{2}{3}u_\mu u_\nu \right)\,.
\label{Gmunu}
\ea 
The metric $G_{\mu\nu}$ above, termed the ``acoustic metric'',
is described by the line element\,\footnote{We will use $G_{\mu\nu}$ and $d\tilde{s}^2$ for the acoustic
metric to distinguish it from the spacetime metric.  Here the
spacetime metric has the form $ds^2 = -dt^2 + g_{ij}dx^idx^j$} \be
d\tilde{s}^2 = \sqrt{3}\TT^2\left\{- (1-\frac{2}{3}\gamma^2)dt^2
-\frac{4}{3}\gamma^2v_idx^idt + \left( g_{ij} +
\frac{2}{3}\gamma^2v_iv_j\right) dx^idx^j \right\}\,,
\label{ds2_acous}
\ee
where $\gamma(z)=1/\sqrt{1-v(z)^2}$ is the Lorentz factor for the moving fluid.
If we make the transformation
$d\tau=dt+\frac{\frac{2}{3}\gamma^2v_i}{1-\frac{2}{3}\gamma^2}dx^i$,
the metric becomes \be
\label{dst2}
d\tilde{s}^2 = \sqrt{3}\TT^2\left\{ -(1-\frac{2}{3}\gamma^2)d\tau^2 +
\left( g_{ij} +
\frac{\frac{2}{3}\gamma^2}{1-\frac{2}{3}\gamma^2}v_iv_j\right)
dx^idx^j \right\}\,. \ee This is the most general form of the acoustic
metric for a relativistic conformal fluid. Note
that the metric factors vanish or become singular at
$\gamma=\sqrt{\frac{3}{2}}$ which precisely corresponds to the
speed of sound $v=c_s=\frac{1}{\sqrt{3}}$.  This indicates that an
``acoustic horizon'' is formed where the flow becomes supersonic 
and sound waves do not emerge from out of that horizon.

\subsection{Steady flows leading to acoustic horizons}
In this section we will find steady fluid flows with acoustic horizons
when the background spacetime has a metric of the form \be
\label{ds2R(z)}
ds^2 = -dt^2 +dz^2 +R(z)^2 d\Omega_2^2\,.  \ee $R(z)=z$ corresponds to flat
spacetime; we will later consider more general functions $R(z)$ and work
on spacetimes that are asymptotically flat.  We also assume that the
thermodynamic quantities depend only on $z$ and that $v_z(z)$ is the only
nonzero component of the velocity.  The form of the acoustic metric
is then \ba
\label{dst2R(z)}
d\tilde{s}^2 &=&
\sqrt{3}\TT^2\left\{-(1-\frac{2}{3}\gamma(z)^2)d\tau^2 +
\frac{dz^2}{3(1-\frac{2}{3}\gamma(z)^2)} + R(z)^2 d\Omega_2^2 \right\}
\\ &=&
\sqrt{3}\TT^2\left\{-c_s^2\,\gamma(z)^2(1-\frac{v_z(z)^2}{c_s^2})d\tau^2
+ \frac{dz^2}{\gamma(z)^2(1-\frac{v_z(z)^2}{c_s^2})} +
R(z)^2d\Omega_2^2 \right\}\,. \nn \ea 
The second line can be obtained from the first by using the definition of the Lorentz factor.
Up to an overall conformal
factor the metric is remarkably similar to that of a Schwarzschild
black hole with a warp factor proportional to $(1-\frac{2}{3}\gamma(z)^2)$ -- an acoustic horizon is present at the radius where the flow
becomes supersonic.  An acoustic Hawking temperature $T_H$ and a
surface gravity $\kappa$ can be defined by the standard process of
Euclidean continuation of the acoustic metric near the horizon; then \be
\label{TH}
T_H = 
\frac{\kappa}{2\pi} = \frac{3}{4\pi}\left|\frac{dv_z}{dz}\right|_{z_h}\,.
\ee 
Thermal radiation of quantized phonons is expected from the horizon since Hawking radiation is a purely kinematic effect independent of the underlying dynamics \cite{Visser:2001kq}.

In order to get an explicit solution, we need to solve the equations of motion (\ref{perp_vort}) and (\ref{entropy})
\ba
\label{T_infty}
\partial_z(\TT\gamma) = 0 &\implies& 
\TT\gamma = \TT_\infty \\
\partial_z\left(R(z)^2\TT^3\gamma v_z\right) = 0 &\implies&
R(z)^2\TT^3\gamma{v_z} = \Phi_S
\label{Phi_S}
\ea 
where we have identified the integration constants as the
``asymptotic temperature'' $\TT_\infty$ and the ``entropy flux''
$\Phi_S$.  From (\ref{T_infty}) and (\ref{Phi_S}) \
\be
\label{v_cubic}
v_z(1-v_z^2) = \frac{\Phi_S}{\TT_\infty^3}\frac{1}{R(z)^2}\,.
\ee 
From the isentropic condition $q\sim\TT^3$ it follows that
\be
\label{q_infty}
q(z) = \frac{q_\infty}{\gamma^3(z)}\,.
\ee

\subsubsection{Singular radially symmetric solution}
If we take $R(z)=z$ with $z\in(0,\infty)$, the metric (\ref{ds2R(z)})
describes flat spacetime. The LHS of Eqn.(\ref{v_cubic}) has a maximum
value of $\frac{2}{3\sqrt3}$ for $v=c_s=\frac{1}{\sqrt{3}}$, and therefore
there is no solution for $z$ below a minimum value of \be
\label{zh_flat}
z_h = \left( \frac{\Phi_S}{\TT_\infty^3}\frac{3\sqrt3}{2} \right)^\frac{1}{2}\,.
\ee
\be
\label{v_flat}
\Rightarrow v_z(1-v_z^2) = \frac{\Phi_S}{\TT_\infty^3}\frac{1}{z^2} = \frac{2}{3\sqrt{3}}\frac{z_h^2}{z^2}\,.
\ee
\begin{figure}
\begin{center}
\scalebox{0.3}{\includegraphics{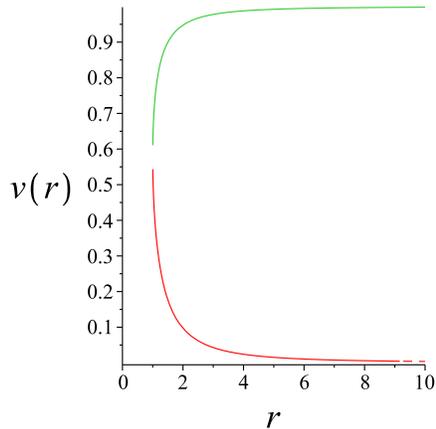}}
\caption{Plot of $v_z(z)$ for the spherically symmetric case given by Eqn.(\ref{v_flat}) with $z_h=1$. There are two physical branches; neither is valid for $z<z_h$. The third branch is superluminal.}
\end{center}
\label{plot_flat}
\end{figure} 
This cubic
equation for $v_z$ has two physical branches, one subsonic
and the other supersonic; the third branch is superluminal.
The physical branches are plotted in Fig.~1.  In
the subsonic branch, as $z\to\infty$, $v_z\to0$
which is consistent with the identification (\ref{T_infty}). 
At the horizon the derivatives blow up:
\be
-\frac{dv_z}{dz} = \frac{\Phi_S}{\TT_\infty^3}\frac{2}{3z^3(1-3v_z^2)} \to \infty {\rm \ at\ } \ z=z_h.
 \label{dv_dz}
\ee
As a result, quantities like the Hawking temperature blow up and more importantly the hydrodynamic description breaks down. The pathology can possibly be cured by introducing a viscosity \cite{Liberati:2000pt}, {\em i.e.\!} by going to higher orders of the hydrodynamic derivative expansion.

\subsubsection{Solution in more general geometries}
To obtain solutions of first order hydrodynamics that are valid globally, 
we can choose a more general $R(z)$ 
such that
\begin{itemize}
\item the spacetime is asymptotically flat, $R(z)\sim |z|$ for $z\to\pm\infty$
\item the maximum value of the RHS of (\ref{v_cubic}) is $\frac{2}{3\sqrt3}$, the same as the maximum possible value of the LHS.  This condition implies that the minimum value attained by $R(z)$ is 
\end{itemize}
\be
\label{R(z)_min}
R_{\rm min}=\left(\frac{3\sqrt3}{2}\frac{\Phi_S}{\TT_\infty^3}\right)^{1/2}.
\ee
Then we can construct a smooth solution that changes over from subsonic to supersonic or vice-versa every time $R(z)$ attains the above minimum value.
For a generic velocity profile, the derivatives at the horizon do not blow up because the divergence of $\frac{dv_z}{dR(z)}$ at the horizon is canceled by the fact that 
\be
\label{dRdz}
\frac{dR(z)}{dz}= -R(z)\frac{1-3v_z^2}{2v_z(1-v_z^2)}\frac{dv_z}{dz} =0 {\rm \ at }\ z=z_h.
\ee
The simplest example is the wormhole geometry given by 
\be
\label{R(z)_wh}
R(z)=\sqrt{z^2+z_0^2}\,.
\ee 
There are two asymptotically flat sheets as $z\to\pm\infty$ which are
connected by a throat at $z=0$, where $R(z)=R_{\rm min}=z_0$.
From (\ref{R(z)_min}) we get the condition $\frac{\Phi_S}{\TT_\infty^3}=\frac{2z_0^2}{3\sqrt3}$, and (\ref{v_cubic}) then gives
\be
\label{v_wh}
v_z(1-v_z^2) = \frac{\Phi_S}{\TT_\infty^3}\frac{1}{R(z)^2} = \frac{2}{3\sqrt{3}}\frac{z_0^2}{z^2+z_0^2}\,.
\ee
The cubic equation for can be solved for $v_z$. Again there are two physical (subluminal) branches. 
One of them smoothly increases from $v_z=0$ to $v_z=1$ for $z\in(-\infty,\infty)$ while the other one smoothly decreases. Both solutions have acoustic horizons at $z=0$. 
A plot of the solutions is given in Fig.~2.
\begin{figure}
\begin{center}
\scalebox{0.3}{\includegraphics{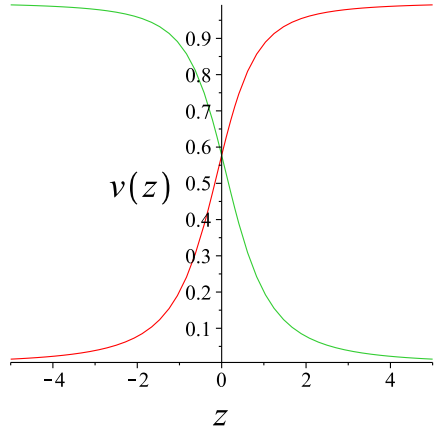}}
\label{plot_wh}
\caption{Plot of $v_z(z)$ for the wormhole geometry (\ref{R(z)_wh}) with $z_0=1$. There are two physical branches, making a subsonic to supersonic (supersonic to subsonic) transition at the acoustic horizon at $z=0$.}
\end{center}
\end{figure}
The velocity and its derivatives remain finite near the horizon, as seen from the near-horizon expansion of the increasing solution:
\be
\label{v_wh_nh}
v_z=\frac{1}{\sqrt3}+\frac{\sqrt2}{3}\frac{z}{z_0}
\ee
and the acoustic Hawking temperature is
\be
\label{TH_wh}
T_H=\frac{3}{4\pi}\left|\frac{dv_z}{dz}\right|_{z=0}=\frac{1}{2\sqrt2\pi z_0}\,.
\ee
An obvious problem with this solution is that it reaches the speed of
light asymptotically on one of the sheets.

To fix this problem, we can choose, for example
\be
\label{R(z)_2wh}
R(z)=(z^4-2z^2z_0^2+z_0^4+R_{\rm min}^4)^{\frac{1}{4}}
\ee
which is smooth for $R_{\rm min}>0$ and has minima at $z=\pm z_0$ where $R(z)=R_{\rm min}$. (\ref{R(z)_min}) sets $\frac{\Phi_S}{\TT_\infty^3}=\frac{2R_{\rm min}^2}{3\sqrt3}$. This is geometry where two asymptotically flat regions separated by a wormhole with two throats. For the velocity profile, we have
\be
\label{v_2wh}
v_z(1-v_z^2) = \frac{\Phi_S}{\TT_\infty^3}\frac{1}{R(z)^2} = \frac{2}{3\sqrt{3}}\frac{R_{\rm min}^2}{\sqrt{z^4-z^2z_0^2+z_0^4+R_{\rm min}^4}}\,.
\ee
Now we can find a solution that crosses over from subsonic to supersonic and back to subsonic at the horizons and remains subluminal for all $z$, as shown in Fig.~3.
\begin{figure}
\begin{center}
\scalebox{0.3}{\includegraphics{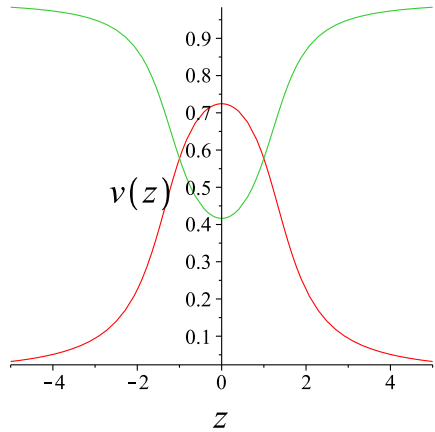}}
\label{plot_2wh}
\caption{Plot of $v_z(z)$ for the geometry with two throats given by Eqn.(\ref{R(z)_2wh}) with $z_0=1$, $R_{\rm min}=1$. There are two horizons located at $z_h=\pm z_0$. The red branch remains subluminal for all $z$.}
\end{center}
\end{figure}

\subsubsection{Nozzle geometry}
\label{ss:nozzle}
We can also choose the spatial section of the metric to be asymptotically cylindrical with $R(z)\to$ constant at $z\to\pm\infty$.   For variety we consider geometries with a toroidal cross-section 
\be
\label{ds2T2}
ds^2 = -dt^2 +dz^2 +R(z)^2 (d\theta_1^2+d\theta_2^2).
\ee
The form of the acoustic metric (\ref{ds2R(z)}) remains very similar with the replacement of $d\Omega_2^2$ by $d\theta_1^2+d\theta_2^2$. The acoustic Hawking temperature is still given by (\ref{TH}). 
As an example, we can take the following profile for $v(z)$ and then solve (\ref{v_cubic}) for $R(z)$: 
\ba
v_z &=& v_{\rm min} + (v_{\rm max}-v_{\rm min})\,\sech\left(\frac{z}{z_0}\right)\,.
\label{explicit_profile}
\ea
The $v_{\rm min}$ term ensures that $R(z)\to$ constant and remains finite for large $|z|$.
The profiles of $v(z)$ and $R(z)$ are shown in Fig.~4.
\begin{figure}
\begin{center}
\scalebox{0.3}{\includegraphics{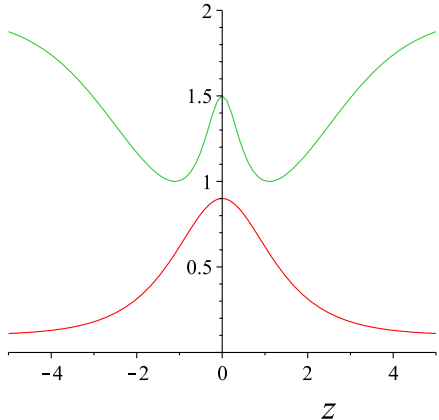}}
\label{plot_nozzle}
\caption{Velocity profile (in red) for the nozzle geometry given by Eqn.(\ref{explicit_profile}) with $v_{\rm min}=0.1$, $v_{\rm max}=0.9$, $z_0=1$ and the corresponding $R(z)/R_{\rm min}$ (in green). The minima in $R(z)$ at $z=\pm z_0$ correspond to $v=c_s$.}
\end{center}
\end{figure}
There are two horizons located at 
\be
\label{horizon_location}
z_\pm = \pm z_0\cosh^{-1}\frac{v_{\rm max}-v_{\rm min}}{c_s-v_{\rm min}}\,.
\ee
Assuming $v_z>0$, the fluid passes the speed of sound at the
left horizon $z_-$ and again returns to subsonic speeds as it crosses the right horizon
$z_+$.

\subsection{Sound waves}

In this subsection we examine the behavior of sound waves around the
background flows described in the previous subsections. Sound waves
are fluctuations of the velocity potential $\phi$ which satisfy a
massless Klein-Gordon equation in the background acoustic metric (\ref{dst2R(z)}).
We will consider background flows on a simple wormhole geometry - 
more complicated wormholes or 
nozzle geometries can be treated along similar lines. 

Near the acoustic horizon (chosen at $z=0$), the metric is similar to
the usual Schwarzschild metric, so we expect a large blueshift effect
for outgoing modes. To display this, it is sufficient to consider
modes in the $s$-wave. Let us first rewrite the metric in terms of null
coordinates $u,v$ as follows
\begin{equation}
\label{line_element_null}
d\tilde{s}^2 = \sqrt3\TT^2(z) \left\{ -(1-\frac{2}{3}\gamma^2)dudv +
R^2(z)(d\theta^2+\sin^2\theta d\phi^2) \right\},
\end{equation}
where
\ba
\label{uv}
&& du= d\tau-\frac{dz}{\sqrt3(1-\frac{2}{3}\gamma^2)} =
dt+\frac{\frac{2}{3}\gamma^2v\,dz}{1-\frac{2}{3}\gamma^2}
-\frac{dz}{\sqrt3(1-\frac{2}{3}\gamma^2)} = dt - \frac{dz}{v_+(z)}
\nn \\ 
&& dv= d\tau+\frac{dz}{\sqrt3(1-\frac{2}{3}\gamma^2)} =
dt+\frac{\frac{2}{3}\gamma^2v\,dz}{1-\frac{2}{3}\gamma^2}
+\frac{dz}{\sqrt3(1-\frac{2}{3}\gamma^2)} = 
dt - \frac{dz}{v_-(z)}
\,.  
\ea 
Here
$v_\pm (z) =\frac{v (z) \pm c_s}{1\pm v (z) c_s}$ is the relativistic sum of
the local fluid velocity and the velocity of sound $c_s = 1/\sqrt{3}$.
Then close to the acoustic horizon $R(z) \rightarrow R(0)$ is
finite, and the $s$-wave solutions of the Klein-Gordon equation are approximately
\be
\label{psi_+-}
\psi_+\sim e^{-i\omega u} \hspace{1cm} \psi_-\sim e^{-i\omega v} \,.
\ee 
In the asymptotic region, where the velocity becomes constant, we
obtain the usual spherical Bessel functions.
To analyze the near-horizon behavior, we can expand
the velocity field near the horizon as
$
v(z) \approx c_s + \frac{2}{3}\kappa z + \ldots
$,
where $\kappa=\frac{3}{2}\left|\frac{dv}{dz}\right|_{z_h}$ is the surface gravity at the horizon.
Using this in (\ref{psi_+-}), we find 
\ba
\label{psiplus_near_horiz}
\psi_+&\sim& e^{-i\omega\left[t-\frac{2z}{\sqrt3}+O(z^2) \right]} \\
\label{psiminus_near_horiz}
\psi_-&\sim&
e^{-i\omega\left[t-\frac{1}{\kappa}\ln\left|z\right|+\frac{z}{\sqrt3}+O(z^2)
\right]} \ea 
The $\psi_+$ mode is continuous at the horizon and is
right moving with a velocity $\frac{\sqrt3}{2}$, which is the
relativistic sum of $\frac{1}{\sqrt3}$ with itself.  The $\psi_-$ mode
has rapid oscillations near the horizon: \ba z\lesssim 0\ : &&
\psi_-\sim e^{-i\omega\left[t-\frac{1}{\kappa}\ln(-z)\right]}
\hspace{0.5cm} \mbox{(left-moving)} \nn \\ z\gtrsim 0\ : && \psi_-\sim
e^{-i\omega\left[t-\frac{1}{\kappa}\ln(z)\right]} \hspace{0.5cm}\mbox {\rm
(right-moving)} \nn
\ea
indicating that inside the horizon, both modes are right-moving.

To extend these modes away from the horizon, we employ 
an eikonal approximation.
We decompose the sonic fluctuation into a rapidly varying phase or
``eikonal'' (here $\lambda\gg1$) times a slowly varying envelope \be
\label{Eikonal_ansatz}
\psi(x^\mu) =A(x^\mu) e^{-i\lambda S(x^\mu)}
\ee
and plug it into the wave equation
$\partial_\mu\left[\sqrt{-G}G^{\mu\nu}\partial_\nu\right]\psi=0$
with the acoustic metric (\ref{ds2_acous}) written out in the original
$t$-$z$ coordinates \ba
\label{metric_un_diagonal}
d\tilde{s}^2 = \TT^2 \left\{-(1-\frac{2}{3}\gamma^2)dt^2 -
\frac{4}{3}\gamma^2v_zdtdz + (1+\frac{2}{3}\gamma^2v_z^2) dz^2 +
R(z)^2 (d\theta^2 +\sin^2\theta d\phi^2) \right\}\,.  \ea We then get a sequence of
differential equations for $S(x^\mu)$ and $A(x^\mu)$ by expanding the 
wave equation order by order in $\lambda$:
\begin{eqnarray}
\label{order1_equation}
O(\lambda^{2})&:& \partial_{\mu}S(x^{\alpha}) \partial^{\mu}S
(x^{\alpha})=0 \\
\label{order2_equation}
O(\lambda)&:& 2\partial_{\mu}S(x^{\alpha}) \partial^{\mu}A(x^{\alpha})
+ A(x^{\alpha})\nabla^{2}S(x^{\alpha})=0 \\ &\vdots& \nn
\end{eqnarray}
The leading equation (\ref{order1_equation}) can be used to solve for
$S(x^\alpha)$. Let us consider $s$-wave solutions independent of
$\theta,\phi$. With the ansatz
\begin{equation}
\lambda S(x^{\alpha})=\omega t - f(z)
\end{equation}
we can solve for the phase,
\begin{equation}
\label{phase}
\lambda S_{\pm}(x^{\alpha})=\omega t + \omega\int
dz\,\frac{\frac{2}{3}\gamma^2v \mp
\frac{1}{\sqrt{3}}}{1-\frac{2}{3}\gamma^{2}} = \omega t - \omega\int
dz\, \frac{1\pm\frac{v}{\sqrt3}}{v\pm\frac{1}{\sqrt3}}
\end{equation}
The momenta of the wavepackets are given by 
derivatives of the eikonal\,\footnote{The leading order equation (\ref{order1_equation}) is thus a
null geodesic equation, $p_{\mu}p^{\mu}=0$ for the phonons.}
$p_{\mu}\equiv-\partial_{\mu}[\lambda S(x^\alpha)]$ giving
\begin{eqnarray}
p_{t}&=& \omega
\nn \\
p_{z}&=&
\omega \, \frac{1\pm vc_s}{v\pm c_s} = \frac{\omega}{v_\pm}\,,
\end{eqnarray} 
where $v_\pm$ has been defined above.  The upper and the
lower signs would normally correspond to right- and left-moving sound modes.  
However, if the fluid (assumed to be right moving) has
a velocity greater than the speed of sound, then both the modes become
right-moving.  
Using the ansatz $A(x^\alpha)=A(z)$ in the subleading
equation (\ref{order2_equation}), a full solution is obtained: 
 \be
\label{psi_pm}
\psi_\pm = \frac{A_0}{\TT R(z)}
e^{-i\omega\left[t-\int\frac{dz}{v\pm(z)}\right]} \ee

\section{Gravity dual of acoustic solution}
\label{bulk}

The fluid-gravity correspondence \cite{Policastro:2001yc,Policastro:2002tn,Bhattacharyya:2008jc,Bhattacharyya:2008xc,Bhattacharyya:2008ji,Erdmenger:2008rm,Banerjee:2008th}
provides a correspondence between
  solutions of certain fluids with classical solutions of suitable
  Einstein-Maxwell equations in a $4+1$ dimensional spacetime with a
  negative cosmological constant.  In our case the fluid is a
  conformal fluid with a global U$(1)$ charge and the higher
  dimensional bulk theory is given by the five-dimensional action 
\ben
  S = \frac{1}{16\pi G}\int d^5x~{\sqrt{-g}} \left[ R + 12
  -F_{AB}F^{AB} - \frac{4\kappa}{3}\epsilon^{EABCD}A_EF_{AB}F_{CD}
  \right]
\label{dual-1}
\een
where $G$ is the five dimensional Newton constant, the indices $A,B$ run from
0 to 4, $A_B$ is a U$(1)$ gauge field, and we have chosen units in
which the cosmological constant is $\Lambda = -6$.  The above action is a consistent truncation
of IIB supergravity for $\kappa = 1/(2\sqrt{3})$. We will, however,
allow arbitrary values of $\kappa$.

A uniform charged black brane solution of this action is given in a boosted reference
frame by
\bea
ds^2 & = & - 2u_\mu dx^\mu dr - r^2 V(r,m,\tq) u_\mu u_\nu dx^\mu dx^\nu +
r^2 P_{\mu\nu} dx^\mu dx^\nu \nn \\
A & = & \frac{{\sqrt{3}}\tq}{2r^2}u_\mu dx^\mu 
\label{dual-2}
\eea
where $u_\mu$ are constant 4-velocities (the indices $\mu,\nu = 0
\cdots 3$) and $P_{\mu\nu} = \eta_{\mu\nu} + u_\mu u_\nu$ is the
spatial projection operator. The function $V(r,m,\tq)$ is given by 
\ben
V(r,m,\tq) = 1 - \frac{m}{r^4}+\frac{\tq^2}{r^6}
\label{dual-2a}
\een
where $m$ and $\tq$ are parameters of the solution and we are using the notation
of \cite{Banerjee:2008th}.

This solution is dual to a charged fluid in equilibrium living on the
flat boundary of the five-dimensional spacetime.  
The fluid is strongly-coupled $\mathcal{N}=4$ SU$(N)$ Yang-Mills theory, viewed 
in a boosted frame with coordinates $x^\mu$.
The temperature $T$, charge density $q$, energy density $\epsilon$
and entropy density $s$  of the fluid are given by \cite{Banerjee:2008th}
\ben
T =  \frac{R_+}{2\pi}\left( 2 - \frac{\tq^2}{R_+^6} \right),\hspace{0.5cm}
q=\sqrt3\alpha\tq, \hspace{0.5cm} \epsilon= 3\alpha m, \hspace{0.5cm} s=4\pi\alpha R_+^3, \hspace{0.5cm} \alpha\equiv\frac{1}{16\pi G}
\label{dual-3}
\een
where $R_+$ denotes the radius of the outer horizon, i.e. the largest
root of the equation $V(r,m,\tq) = 0$. The energy momentum tensor
$T_{\mu\nu}$ and the charge current $J_\mu$ of the
fluid are given by
\ben
T_{\mu\nu}  =  \frac{\epsilon}{3}(\eta_{\mu\nu} + 4 u_\mu u_\nu) 
~~~~~~~~~~~~~~~J_\mu  =  q u_\mu
\label{dual-4}
\een
The expressions (\ref{dual-3}) and (\ref{dual-4}) involve the bulk
parameter $G$. In our units, $G$ is related to the rank 
of the gauge group of the boundary theory by
\ben
G = \frac{\pi}{2N^2}, \hspace{1cm} \alpha=\frac{1}{16\pi G} = \frac{N^2}{8\pi^2}.
\label{dual-5}
\een
With the substitutions in (\ref{dual-3}), the equation of state $\epsilon(s,q)$ becomes identical to the condition $V(R_+(s),m(\epsilon),\tq(q))=0$. 
The equation of state for a charged conformal fluid with a gravity dual is thus
\ba
\label{eos_R}
&& 1 - \frac{\epsilon}{3\alpha R_+^4} + \frac{q^2}{3\alpha^2R_+^6} =0, {\rm{\ with\ }} R_+=\left(\frac{s}{4\pi\alpha}\right)^\frac{1}{3} \\
\label{eos_s}
\Rightarrow && \epsilon(s,q)=3\alpha\left(\frac{s}{4\pi\alpha}\right)^\frac{4}{3}+\frac{q^2}{\alpha}\left(\frac{4\pi\alpha}{s}\right)^{\frac{2}{3}}
\ea
The temperature and chemical potential can be obtained by taking derivatives of $\epsilon(s,q)$ using (\ref{T_mu}):
\bea
T & = & \frac{1}{\pi} \left( \frac{s}{4\pi\alpha} \right)^{1/3} 
\left(1 - \frac{8\pi^2 q^2}{3s^2} \right) \nn \\
\mu & = & 2 \left( \frac{q}{\alpha} \right) \left( \frac{4\pi\alpha}{s}
\right)^{2/3}
\label{tmurelations}
\eea
This temperature reproduces the value quoted in (\ref{dual-3}).
In the uncharged limit $R_+=\pi T$, and one can fix the value of $c$ defined in (\ref{ttandrr}) by comparing it with (\ref{eos_R}) and requiring that $\TT=T$ for uncharged fluids; this gives
$c=\alpha\pi^4$ and hence
\ben
\epsilon_{\mu=0} =  3\alpha(\pi T)^4\,,~~~~s_{\mu=0}=
4\pi\alpha(\pi T)^3\,.
\label{unchargedrelations}
\een
For charged fluids at finite $\mu$ there is a zero-temperature limit, 
reached when $R_+=\mu/(2\sqrt6)$ and $\TT=\mu/(192^\frac{1}{4}\pi)$.
In this limit,
\ben
q_{T=0} = \frac{\alpha}{48} \mu^3\,,~~~~
\epsilon_{T=0} = \frac{\alpha}{64} \mu^4\,,~~~~s_{T=0}=
\frac{\pi\alpha}{12\sqrt6}\mu^3\,.
\label{extremalrelations}
\een

As discussed in Section 2 we  restrict our attention to {\em isentropic}
flows. For such flows $q/\TT^3$ is constant and since $\epsilon =
3\alpha\pi^4\TT^4$, it follows from the equation of state
(\ref{eos_R}) that $\TT/R_+$ is a constant, and therefore that
for such flows the entropy per unit charge $s/q$ is constant.
The first equation in (\ref{tmurelations}) shows that by choosing  
\ben
\label{lowtemp}
1- \frac{8\pi^2}{3} \left(\frac{s}{q} \right)^2 \ll 1
\een
we can keep the temperature $T \ll R_+$ everywhere and at all times.

The gravity dual of a general fluid motion is then 
constructed in a derivative expansion as follows. First, we replace
the parameters of the solution by functions of the boundary
coordinates $x^\mu$, $u^\mu \rightarrow u^\mu (x), m \rightarrow m(x),
\tq \rightarrow \tq(x)$ which  respectively  represent the velocity field, energy
density field and the charge density field of the fluid. We also 
replace the flat boundary metric $\eta_{\mu\nu}$ with a curved metric
$g_{\mu\nu} (x)$. With these replacements, (\ref{dual-2}) is no longer a
solution of the bulk equations of motion. Second we need to add
correction terms to the metric and the gauge field so that the full
metric and the gauge field now solve the equations of motion. This
second step is of course impossible to perform in an exact
fashion. However, these corrections can be calculated systematically
in a {\em derivative expansion}, provided that  the
derivatives of $u^\mu(x),\ m(x),\ \tq(x)$ with respect to $x^\nu$ are small
compared to the outer horizon radius $R_+$. To lowest nontrivial
order in the derivative expansion, the modified metric and gauge
fields are
\ba
ds^2&=&-2u_{\mu}dx^{\mu}dr
-r^{2}V(r,m,\tq) u_{\mu}u_{\nu}dx^{\mu}dx^{\nu} + r^{2}
P_{\mu\nu}dx^{\mu}dx^{\nu} \nn \\ \nonumber &+& \frac{2}{3} r
(\nabla_{\alpha} u^{\alpha}) u_{\mu}u_{\nu}dx^{\mu}dx^{\nu} +
\frac{2r^{2}}{R_+} \sigma_{\mu\nu} F_{2}(\rho,M) dx^{\mu}dx^{\nu} \nn \\
&-& 2r u_{\mu}
u^{\alpha}(\nabla_{\alpha}u_{\nu})dx^{\mu}dx^{\nu} \nn \\ 
&-&
2u_{\mu}\left(\frac{\sqrt{3}\kappa \tq^{3}}{m r^{4}}l_{\nu}
+\frac{6 r^{2}}{R_+^7} 
(P^\lambda_\nu \partial_\lambda \tq + 3(u^\lambda \nabla_\lambda u_\nu) \tq)
F_{1}(\rho,M)\right)dx^{\mu}dx^{\nu},
\label{dual-7}
\ea
\ben
A = \left[\frac{\sqrt{3}\tq}{2r^2}u_\mu + \frac{3\kappa \tq^2}{2mr^2}
  l_\mu - \frac{\sqrt{3} r^5}{2 R_+^8}
(P^\lambda_\mu \partial_\lambda \tq + 3(u^\lambda \nabla_\lambda u_\mu)
  \tq) \right] dx^\mu
\label{dual-8} 
\een
where we have defined the quantities
\ben
M = \frac{m}{R_+^4}~~~~~~~~Q =
\frac{\tq}{R_+^3}~~~~~~~\rho = \frac{r}{R_+}\,.
\label{MandQ}
\een
$\nabla_{\mu}$ is a
covariant derivative with boundary metric $g_{\mu\nu}$, and
\ben
l_{\mu}= g_{\mu\gamma}\epsilon^{\nu\alpha\beta\gamma} u_{\nu}
\nabla_{\alpha}u_{\beta}\,, 
~~~~~~~~\sigma_{\mu\nu}=\frac{1}{2}P^{\mu\alpha}P^{\nu\beta}
\left(\nabla_\alpha u_\beta + \nabla_\beta u_\alpha \right) -\frac{1}{3} P^{\mu\nu}(\nabla_{\alpha}
u^{\alpha})\,.
\label{dual-8a}
\een
The functions $F_1(\rho, M,Q)$ and $F_2(\rho,M)$ are defined as
\ba
\label{dual-9}
F_{1}(\rho,M,Q)&=&\frac{1}{3}\left(1-\frac{M}{\rho^4}+\frac{Q^2}{\rho^6}\right)
\int^{\infty}_{\rho} dp
\frac{1}{(1-\frac{M}{\rho^4}+\frac{Q^2}{\rho^6})^{2}}
\left(\frac{1}{p^8}-\frac{3}{4p^7}(1+\frac{1}{M})\right), \\
F_2(\rho,M) &=& \int_\rho^\infty dp \frac{p(p^2+p+1)}{(p+1)(p^4+p^2-M+1)}
\ea
Note that in the above expressions $m,\tq,R_+,M,Q,\rho$ are also
functions of the boundary coordinates $x^\mu$ since $m$ and $\tq$ are
functions of $x^\mu$.

This is a solution of the bulk equations of motion, provided that $m(x)$,
$\tq(x)$ and $u_\mu (x)$ are such that the energy momentum tensor and
current 
\ben 
T_{\mu\nu} = \frac{m(x)}{16\pi G}(g_{\mu\nu}(x) + 4 u_\mu(x) u_\nu
(x))~~~~~
J_\mu = \frac{\sqrt{3}\tq(x)}{16\pi G}  u_\mu (x)
\label{dual-11}
\een
are covariantly conserved,
\ben
\nabla_\mu T^{\mu\nu} = \nabla_\mu J^\mu = 0
\label{dual-12}
\een
Thus every solution of fluid dynamics leads to a bulk solution.

\subsection{Gravity duals of dumb holes}

We now apply the results of the preceding subsection to 
construct gravitational duals of the fluid flows with 
acoustic horizons that were studied in Section 2.  These flows
are special in several ways:  first, the
background spacetime metric of the fluid is of the form (\ref{ds2R(z)}) or
(\ref{ds2T2}) where the only inhomogeneity is in the $z$
direction. Second, both the background flow and the sound wave
fluctuations have vanishing vorticity.  Third, the background flows
as well perturbations around them are isentropic.

It follows from the isentropic condition that the quantities $M$ and $Q$ are
constants. As argued above (see discussion following equation
(\ref{TTform})), for isentropic flows there is just one length
scale, and all quantities are related to this length scale by
dimensional analysis. In particular, the dimensionless quantities $M$ and $Q$ must be
constant.  In addition, the inhomogeneous parts of all quantities which
appear in the bulk metric and the gauge field are determined in terms of a single scalar
field $\phi (x)$.

As discussed in the introduction, in order for the acoustic Hawking radiation to be detectable we 
need to consider fluids which have a very small ambient
temperature. This means that the constant quantities $Q$ and $M$ need to be
close to their extremal values, $Q \approx {\sqrt{2}}$ and therefore
$M \approx 3$. While the various quantities like $\epsilon (x), q(x)$
can change by ${\cal O}(1)$ amounts (only their derivatives are small), the
isentropic condition ensures that if the fluid temperature is initially
small it will remain small (see the discussion in Section 2
above). 

To construct the background fluid flow, we simply need to insert the
velocity potential $\phi_0$ for the solutions of Section 2 into the
general metric and gauge field in (\ref{dual-7}) and
(\ref{dual-8}). The conditions of vanishing vorticity and isentropic
flow simplify these general expressions somewhat. The most drastic
simplification appears in the expression for the gauge field, equation
(\ref{dual-8}). In fact the first order corrections (in the derivative
expansion) to the gauge field vanish for isentropic gradient flows.
To see this we note first that $\nabla_\alpha u_\beta$ can be replaced by
$\partial_\alpha u_\beta$ in the expression $l_\mu$ of
(\ref{dual-8a}). Then using $\TT u_\mu = \partial_\mu \phi$ we get
\ben
l_\mu = g_{\mu\gamma}\epsilon^{\nu\alpha\beta\gamma}\frac{\partial_\nu
  \phi}{\TT}[-\frac{1}{\TT^2} \partial_\alpha \TT \partial_\beta \phi +
  \frac{1}{\TT}\partial_\alpha\partial_\beta \phi ] = 0
\een
due to antisymmetry of the epsilon symbol.  The third term on the RHS of
(\ref{dual-8}) also vanishes, as can be seen by applying the 
isentropic condition $q / \TT^3=$constant  and the relations (\ref{potential}) and
(\ref{TTform}) to the expression 
\bea
 P^\lambda_\mu \nabla_\lambda \tq + 3 (u^\lambda
\nabla_\lambda u_\mu) \tq \nn 
& = & 3a \TT \left[ \TT \partial_\mu \TT + \TT u^\lambda \nabla_\lambda (\TT
  u_\mu) \right] \nn \\
& = & 3a \TT \left[ -\frac{1}{2} \partial_\mu (\partial_\alpha \phi
  \partial^\alpha \phi) + \partial^\lambda \phi \nabla_\lambda \partial_\mu
  \phi \right] = 0
\label{dual-15}
\eea
Thus, to first order in
the derivative expansion, the bulk gauge field is given by the first
term of the right hand side of (\ref{dual-8}), which is just the
term which would have resulted from a simple boost of the original black brane
solution.  In our case the charge density and the 4-velocity appearing in (\ref{dual-8})
are functions of $z$, as determined by the fluid flow on the
boundary. So there is a nonzero electric field component along the $z$
direction, given by
\ben
F_{0z} = -\frac{\sqrt{3} \tq_\infty}{r^2} v_z \partial_z v_z
\label{dual-14}
\een
where we have used (\ref{q_infty}) to express $q(z)$ in terms of the
velocity $v_z$. 

The expression for the bulk metric simplifies as well. In the fourth
line of (\ref{dual-7}), the first term is proportional to $l_\mu$
which vanishes for our flows. The second term is proportional to
$H_\mu$ defined in (\ref{dual-15}) and vanishes as shown above.

In the derivative expansion, the relationship between the boundary and
the bulk becomes essentially local. The bulk solution can in fact be
constructed approximately by patching together tube geometries obtained by
extending the boundary data in a given region of the boundary to the
bulk using the radial equations of motion. Consequently we expect that
the acoustic horizon of the fluid flow on the boundary extends
trivially into the bulk. We will explicitly verify this in the next
subsection.

However, this tubewise approximation breaks down in regions where the
local geometry is exactly extremal. This is apparent in the results of
\cite{Erdmenger:2008rm} and \cite{Banerjee:2008th}. Furthermore,
recent work on perturbations around extremal black holes shows that the
relevant low energy expansion is different from the naive derivative
expansion \cite{Liu:2009dm}. Our results hold close to extremality,
but not exactly at extremality.

\subsection{Gravity duals of phonons}

The gravity duals of phonons in the fluid are quasinormal modes of
metric and gauge field perturbations. Once again, construction of
these modes is trivial. We need to write
\ben
\phi (x^\mu) = \phi_0 (x^\mu) + \beta \delta \phi (x^\mu),
\een
compute $\TT (x)$ and hence $m(x), q(x), u_\mu (x)$ in terms of $\delta\phi$, substitute into
(\ref{dual-7}) and (\ref{dual-8}), and consider the terms which are
linear in $\beta$. By construction, these modes satisfy ingoing
boundary conditions at the bulk horizon. 

The fluctuations of the gauge field $A_\mu$ obtained by this procedure have
a particularly simple form
\ben
\delta A_\mu = -\beta \frac{\sqrt{3} a}{2 r^2} [(\partial_\beta\phi_0)(\partial^\beta\phi_0) \delta^\alpha_\mu +
    (\partial_\mu \phi_0)(\partial^\alpha \phi_0) ]\partial_\alpha
  (\delta \phi) 
\label{dual-16}
\een
For the background flows considered in Section 2, we have found  solutions to the wave equation (\ref{wave_eqn}) for $\delta \phi$  in the region close to the acoustic horizon. 
Upon inserting these solutions into (\ref{dual-16}), we see that the fluctuations
$\delta A_\mu$ have a characteristic behavior near the acoustic
horizon, viz. ingoing waves are smooth while outgoing waves have rapid
fluctuations. This is the precise sense in which the fluctuations 
perceive the acoustic horizon, which has now extended into the
bulk. From the nature of the solution that the extension of the
acoustic horizon into the bulk is rather trivial - i.e. the horizon
perceived by these modes is at the same value of $z$ as the acoustic
horizon on the boundary, and for all values of $r$. 

The fluctuations for the components of the metric can be similarly
worked out and also see a horizon structure at the same value of $z$. 
We therefore conclude that
there are certain quasinormal modes of the bulk metric and the gauge
field which perceive a horizon. If these bulk modes are quantized, one
should find a thermal bath of such modes characterized by the
temperature of the acoustic horizon on the boundary. 

\section{Regime of validity}
\label{regime_validity}

It is important to check that the fluid flow described above is
consistent with the standard conditions for validity of
hydrodynamics. Roughly speaking, hydrodynamics is valid when the
gradients of velocities, temperature and charge densities are small
compared to the inverse mean free path $l_m$. For charged conformal fluids
considered above, there are two scales - the temperature $T$ and the
chemical potential $\mu \equiv \nu T$, so that $ l_m  \sim f(\nu)/T $. The function
$f(\nu)$ is of order one for generic values of $\nu$, but there is an
upper bound on $\nu$, $\nu_c$ where $f(\nu)$ has a simple zero. It is
possible to take 
the limit of $\nu \rightarrow \nu_c$ simultaneously with $T
\rightarrow 0$ such that $l_m$ is finite - the dual of this is in fact
the extremal black hole. 
For the flow described in the previous section, in this limit we have
\bea
\left| \frac{dv_z(z)}{dz} \right| & \ll & \frac{1}{l_m} \nn \\
\left| \frac{1}{\TT} \frac{d \TT (z)}{dz} \right| & \ll & \frac{1}{l_m}
\label{valid-1}
\eea
In particular, since the acoustic Hawking temperature $T_H$ is $\frac{3}{4\pi}\left|\frac{dv_z}{dz}\right|_{z_\mp}$, this implies that
\be
T_H \ll \frac{1}{l_m}
\ee
For observable acoustic Hawking radiation, 
the Hawking temperature should be higher than the fluid temperature. 
So we require 
\ben
T\lesssim T_H
\label{TTH}
\een
Furthermore, the frequency of the sound waves should also be small
compared to the basic scale,  $\omega\ll 1/l_m$. However, the finite
Hawking temperature is going to introduce an upper bound on the
allowed wavelengths, due to periodicity in Euclidean time; thus
$\omega>T_H$. Thus we need  
\be
\label{regime}
T \lesssim T_H < \omega \ll \frac{1}{l_m}
\ee
For fluids with no conserved charge, there is only one energy scale, namely, the temperature $T$; thus $1/l_m \sim T$ and (\ref{TTH}) cannot be satisfied.
Although the solution is otherwise valid, the Hawking radiation, at a temperature much
lower than the ambient temperature, is not going to be observable. 
For fluids with a conserved charge the condition (\ref{TTH}) does not pose a problem because 
now we have two length scales, the temperature $T$ and the chemical potential $\mu$. 
For fluids very close to zero temperature, the mean free path will be governed only by $\mu$. We can thus have
\be
0 \approx T \lesssim T_H < \omega \ll \frac{1}{l_m} \approx \TT \approx \mu
\ee 

The ability to construct a gravity dual using a derivative expansion
imposes further conditions. 
In the presence of a
nonconstant $R(z)$, the validity of the derivative expansion of the solutions of
the bulk equations of motion requires that the curvature invariants and all invariants constructed out of the derivatives of curvature be small. An example of such an invariant is $g^{\mu\nu}\nabla_\mu \mcR\nabla_\nu\mcR$, where $\mcR$ is the Ricci scalar and we require for this example
\be
\label{deriv_curv}
(g^{\mu\nu}\,\nabla_\mu\,\mcR\nabla_\nu\,\mcR)^{\frac{1}{6}}\ll\frac{1}{l_m}\,.
\ee

We get  additional conditions if some of
the boundary directions are compactified as in the nozzle geometry of Section (\ref{ss:nozzle}). 
If one boundary direction of a $AdS
\times S$ geometry is made compact with a radius $R(z)$, 
the dual is an $AdS$ soliton \cite{Horowitz:1998ha} which
caps off the geometry at a value of the radial coordinate $r =
1/(2R)$. For a black brane geometry, compactification of a boundary
direction would lead to a similar modification of the usual black
brane geometry. However if $R(z)\gg 1/(2R_+)$, where $R_+$ is the
location of the black brane horizon, the place where the bulk
geometry would cap off is far inside the black brane horizon. In this
situation we can continue to use the standard black brane geometry
with a compact longitudinal direction. We will therefore require that
for all $z$, 
\ben
R(z) \gg 1/R_+
\label{valid-2}
\een
for the solution of Section (\ref{ss:nozzle}).
Finally we require for the nozzle solution that $R(z)$ be finite for large $z$. 
The geometry is then asymptotically $\mathbb{R}\times{T}^2$ and has an AdS dual.

\subsection{Validity of our solutions}

Finally, we determine the range of parameters for which our approximations are valid, for the specific flows studied in Section 2. 
Let us first discuss the wormhole solution of equations (\ref{R(z)_2wh}) and (\ref{v_2wh}).
The solution has four parameters, $\TT_\infty$, $q_\infty$, $z_0$ and $R_{\rm min}$.
$\Phi_S$ is fixed by (\ref{R(z)_min}) once $R_{\rm min}$ is chosen.
Since $v<1$, $\gamma$ remains finite for all $z$. $\TT=\frac{\TT_\infty}{\gamma}>0$ and we can have a valid derivative expansion w.r.t. $\TT_\infty$.
In order that the curvatures are small, we require $R_{\rm min}\gg1/\TT_\infty$. 
The derivatives $\frac{dv_z}{dz}$, $\frac{1}{\TT}\frac{d\TT}{dz}$ and the derivatives of the curvature are all proportional to $\frac{1}{z_0}$. Thus we require $\frac{1}{z_0}\ll\TT_\infty$.
$q_\infty$ can be chosen such that we are always at very low temperatures, following the discussion around equation (\ref{lowtemp}).
To summarize, the conditions for our wormhole solution to be valid are:
\be
\label{valid_2wh}
T\to0,\hspace{0.5cm}\frac{1}{z_0}<\omega\ll\TT_\infty\,,\hspace{0.5cm}\frac{1}{R_{\rm min}}\ll\TT_\infty.
\ee

For the nozzle solution described by (\ref{explicit_profile}), the parameters are $z_0$, $v_{\rm max}$, $v_{\rm min}$, $\TT_\infty$, $q_\infty$ and $\Phi_S$.
As in the previous case, all derivatives in the solution are proportional to $\frac{1}{z_0}$.  We need $v_{\rm max}<1$ for the derivative expansion to be valid (so that $\gamma$ remains finite and $\TT$ remains non-zero) and we obtain the same conditions as in (\ref{valid_2wh}). The condition (\ref{valid-2}) is same as the requirement that the background curvature remains small: $R_{\rm min}\gg1/\TT_\infty$. Since $R_{\rm min}$ is given in terms of $\TT_\infty$ and $\Phi_S$ by (\ref{R(z)_min}), this implies $\Phi_S\gg \TT_\infty$.
Moreover, we need $v_{\rm min}>0$ in order that $R(z)$ be finite at large $z$ -- the asymptotic geometry remains $\mathbb{R}\times{T}^2$, and we have an asymptotically AdS gravity dual. Summarizing, the conditions for validity of our nozzle solution are:
\be
T\to0,\hspace{0.5cm}\frac{1}{z_0}<\omega\ll \TT_\infty \ll \Phi_S,\hspace{0.5cm} 0< v_{\rm min} < v_{\rm max}<1.
\ee

\section*{Acknowledgements}
We would like to thank Allan Adams, Sayantani Bhattacharyya, Jyotirmoy
Bhattacharya, Ian Ellwood, Dileep Jatkar, R.~Loganayagam, Oleg Lunin, Gautam Mandal,
Shiraz Minwalla and Aninda Sinha for
discussions. This work was partially supported by a National Science
Foundation grant NSF-PHY-0855614.

\appendix

\end{document}